\footnotesep \baselineskip	
\everypar={\hangafter=1 \hangindent=3.5em \parindent=0pt \relax}
{ \normalbaselineskip=8pt \parskip 0pt \parindent=0pt
\everypar={\hangafter=1 \hangindent=3.5em \parindent=0pt \relax}
\small
{\bf figure~ #1\/:}
}
{}
\newcommand{\ax}{\ifmmode{\alpha_x} \else $\alpha_x$\fi} 
\newcommand{\ar}{\ifmmode{\alpha_r} \else $\alpha_x$\fi} 
\newcommand{\aE}{\ifmmode{\alpha_E} \else $\alpha_E$\fi} 
\newcommand{\atoms}{\ifmmode{\rm ~atoms~cm^{-2}} \else ~atoms cm$^{-2}$\fi}
\newcommand{\nh}{\ifmmode{\rm N_{H}} \else N$_{H}$\fi}
\newcommand{\nhgal}{\ifmmode{ N_{H}^{Gal}} \else N$_{H}^{Gal}$\fi}
\newcommand{\nhintr}{\ifmmode{ N_{H}^{intr}} \else N$_{H}^{intr}$\fi}
\newcommand{\nhtot}{\ifmmode{ N_{H}^{tot}} \else N$_{H}^{tot}$\fi}
\newcommand{\fx}{\ifmmode l_x \else $~f_x$\fi}
\newcommand{\logfx}{\ifmmode{\rm log}~f_x \else log$~f_x$\fi}
\newcommand{\lopt}{\ifmmode l_{opt} \else $~l_{opt}$\fi}
\newcommand{\loglopt}{\ifmmode{\rm log}~l_{opt} \else log$~l_{opt}$\fi}
\newcommand{\lx}{\ifmmode l_x \else $~l_x$\fi}
\newcommand{\loglx}{\ifmmode{\rm log}~l_x \else log$~l_x$\fi}
\newcommand{\aox}{\ifmmode{\alpha_{ox}} \else $\alpha_{ox}$\fi} 
\newcommand{\aro}{\ifmmode{\alpha_{ro}} \else $\alpha_{ro}$\fi} 
\newcommand{\auv}{\ifmmode{\alpha_{uv}} \else $\alpha_{uv}$\fi} 
\newcommand{\auvx}{\ifmmode{\alpha_{uvx}} \else $\alpha_{uvx}$\fi} 
\newcommand{\luv}{\ifmmode l_{uv} \else $~l_{uv}$\fi}
\newcommand{\logluv}{\ifmmode{\rm log}\,l_{uv} \else log$\,l_{uv}$\fi}
\newcommand{\bmv}{\ifmmode{(B-V)} \else $(B-V) \fi}
\newcommand{\ebmv}{\ifmmode{\rm E}_{B-V} \else E$_{B-V}$\fi}
\newcommand{\ein}{{\em Einstein{\rm}}}

\newcommand{\ros}{{\em ROSAT{\rm}}}
\newcommand{\IUE}{{\em IUE}}

\newcommand{\flam}{\ifmmode f_{\lambda} \else$f_{\lambda}$\fi}
\newcommand{\nufnu}{\ifmmode \nu f_{\nu} \else$\nu f_{\nu}$\fi}
\newcommand{\fnu}{\ifmmode f_{\nu} \else$f_{\nu}$\fi}
\newcommand{\fcgs}{\ifmmode erg~cm^{-2}~s^{-1[B}\else erg~cm$^{-2}$~s$^{-1}$\fi}
\newcommand{\lcgs}{\ifmmode erg~~s^{-1[B}\else erg~s$^{-1}$\fi}
\newcommand{\flamcgs}{\ifmmode erg~cm^{-2}~s^{-1}~Hz^{-1}\else erg~cm$^{-2}$~s$^{-1}~$\AA$^{-1}$\fi}
\newcommand{\fnucgs}{\ifmmode erg~cm^{-2}~s^{-1}~Hz^{-1}\else erg~cm$^{-2}$~s$^{-1}$~Hz$^{-1}$\fi}
\newcommand{\lnucgs}{\ifmmode erg~s^{-1}~Hz^{-1}\else erg~s$^{-1}$~Hz$^{-1}$\fi}
\newcommand{\kms}{\ifmmode~{\rm km~s}^{-1}\else ~km~s$^{-1}~$\fi}
\newcommand{\mone}{\ifmmode ^{-1}\else$^{-1}$\fi}
\newcommand{\mtwo}{\ifmmode ^{-2}\else$^{-2}$\fi}
\newcommand{\arcsec}{\ifmmode ^{\prime\prime}\else$^{\prime\prime}$\fi}
\newcommand{\arcmin}{\ifmmode ^{\prime}\else$^{\prime}$\fi}
\newcommand{\degs}{\ifmmode ^{\circ}\else$^{\circ}$\fi}
\newcommand{\mv}{\ifmmode {m_{V}}\else${m_{V}}$\fi}
\newcommand{\Mv}{\ifmmode {M_{V}}\else${M_{V}}$\fi}
\newcommand{\lapprox}{<\atop^\sim}  
\newcommand{\aliii}{\ifmmode{{\rm Al\,III}} \else Al\,III\fi}
\newcommand{\ciii}{\ifmmode{{\rm C\,III]}} \else C\,III]\fi}
\newcommand{\civ}{\ifmmode{{\rm C\,IV}} \else C\,IV\fi}
\newcommand{\hei}{\ifmmode{{\rm He\,I}} \else He\,I\fi}
\newcommand{\heii}{\ifmmode{{\rm He\,II}} \else He\,II\fi}
\newcommand{\lya}{\ifmmode{{\rm Ly}\alpha}\else Ly$\alpha$\fi}
\newcommand\lyb{\ifmmode {\rm Ly}\beta \else Ly$\beta$\fi}
\newcommand{\mgii}{\ifmmode{{\rm Mg\,II}} \else Mg\,II\fi}
\newcommand{\nv}{\ifmmode{{\rm N\,V}} \else N\,V\fi}
\newcommand{\feii}{\ifmmode{{\rm Fe\,II}} \else Fe\,II\fi}
\newcommand{\ovi}{\ifmmode{{\rm O\,VI}} \else O\,VI\fi}
\newcommand{\siiv}{\ifmmode{{\rm Si\,IV}} \else S\,IV\fi}
\newcommand\hb{\ifmmode {\rm H}\beta \else H$\beta$\fi}
\newcommand\hg{\ifmmode {\rm H}\gamma \else H$\gamma$\fi}
\newcommand\hd{\ifmmode {\rm H}\delta \else H$\delta$\fi}
\newcommand{\oiii}{\ifmmode{\rm [O\,III]} \else [O\,III]\fi}
\newcommand{\ew}{\ifmmode{W_{\lambda}} \else $W_{\lambda}$\fi}
\newcommand{\wciii}{\ifmmode{W_{\lambda}({\rm C\,III]})} \else $W_{\lambda}$(C\,III])\fi}
\newcommand{\wciv}{\ifmmode{W_{\lambda}({\rm C\,IV})} \else $W_{\lambda}$(C\,IV)\fi}
\newcommand{\wfeii}{\ifmmode{W_{\lambda}({\rm Fe\,II})} \else $W_{\lambda}$(Fe\,II)\fi}
\newcommand{\wheii}{\ifmmode{W_{\lambda}({\rm He\,II})} \else $W_{\lambda}$(He\,II)\fi}
\newcommand{\wlya}{\ifmmode{W_{\lambda}({\rm Ly}\alpha)}\else $W_{\lambda}$(Ly$\alpha$)\fi}
\newcommand\wlyb{\ifmmode{ W_{\lambda}({\rm Ly}\beta )} \else $W_{\lambda}$(Ly$\beta$)\fi}
\newcommand{\wmgii}{\ifmmode{W_{\lambda}({\rm Mg\,II})} \else $W_{\lambda}$(Mg\,II)\fi}
\newcommand{\wovi}{\ifmmode{W_{\lambda}({\rm O\,VI})} \else $W_{\lambda}$(O\,VI)\fi}
\newcommand{\wsiiv}{\ifmmode{W_{\lambda}({\rm Si\,IV})} \else $W_{\lambda}$(Si\,IV)\fi}
\newcommand\whb{\ifmmode{ W_{\lambda}({\rm H}\beta )} \else $W_{\lambda}$(H$\beta$)\fi}
\newcommand\whg{\ifmmode{ W_{\lambda}({\rm H}\gamma )} \else $W_{\lambda}$(H$\gamma$)\fi}
\newcommand\whd{\ifmmode{ W_{\lambda}({\rm H}\delta )} \else $W_{\lambda}$(H$\delta$)\fi}
\newcommand{\woii}{\ifmmode{{W_{\lambda}(\rm [O\,II])}} \else $W_{\lambda}$([O\,II])\fi}
\newcommand{\woiii}{\ifmmode{{W_{\lambda}(\rm [O\,III])}} \else $W_{\lambda}$([O\,III])\fi}
\newcommand{\rciii}{\ifmmode{{\rm C\,III]/Ly} \alpha} \else C\,III]/Ly$\alpha$\fi}
\newcommand{\rciiiciv}{\ifmmode{{\rm C\,III]/C\,IV}} \else C\,III]/C\,IV\fi}
\newcommand{\rciv}{\ifmmode{{\rm C\,IV/Ly} \alpha} \else C\,IV/Ly$\alpha$\fi}
\newcommand{\rheii}{\ifmmode{{\rm He\,II/Ly} \alpha} \else He\,II/Ly$\alpha$\fi}
\newcommand{\rmgii}{\ifmmode{{\rm Mg\,II/Ly} \alpha} \else Mg\,II/Ly$\alpha$\fi}
\newcommand{\rovi}{\ifmmode{{\rm O\,VI/Ly} \alpha} \else O\,VI/Ly$\alpha$\fi}
\newcommand{\roviciv}{\ifmmode{{\rm O\,VI/C\,IV}} \else O\,VI/C\,IV\fi}
\newcommand{\rsiiv}{\ifmmode{{\rm Si\,IV/Ly} \alpha} \else Si\,IV/Ly$\alpha$\fi}
\newcommand{\roiii}{\ifmmode{{\rm [O\,III]/H} \beta} \else [O\,III]/H$\beta$\fi}
\newcommand{\neiii}{\ifmmode{\rm [Ne\,III]} \else [Ne\,III]\fi}
\newcommand{\wneiii}{\ifmmode{{W_{\lambda}(\rm [Ne\,III])}} \else \$W_{\lambda}($[Ne\,III])\fi}
\newcommand{\oii}{\ifmmode{\rm [O\,II]} \else [O\,II]\fi}
\newcommand{\nev}{\ifmmode{\rm [Ne\,V]} \else [Ne\,V]\fi}
\newcommand{\wnev}{\ifmmode{{W_{\lambda}(\rm [Ne\,V])}} \else \$W_{\lambda}($[Ne\,V])\fi}
\newcommand{\cii}{\ifmmode{\rm C\,II]} \else C\,II]\fi}
\newcommand{\wcii}{\ifmmode{{W_{\lambda}(\rm C\,II])}} \else \$W_{\lambda}($C\,II])\fi}
\newcommand{\sii}{\ifmmode{{\rm [Si\,II]}} \else [S\,II]\fi}
\newcommand{\wsii}{\ifmmode{W_{\lambda}({\rm [Si\,II]})} \else $W_{\lambda}$([Si\,II])\fi}

\documentstyle[epsf,emulateapj]{article}

\lefthead{Green, P. J.}
\righthead{SPECTRA OF X-RAY BRIGHT AND X-RAY FAINT QSOS}

\begin{document}
\received{July 29, 1997}
\accepted{December 4, 1997}

\title{ Differences Between the Optical/UV Spectra of 
            X-ray Bright and X-ray Faint QSOs}

\author{Paul J. Green\altaffilmark{1} }

\affil{Harvard-Smithsonian Center for
Astrophysics, 60 Garden St., Cambridge, MA 02138}
\altaffiltext{1}{pgreen@cfa.harvard.edu}

\begin{abstract}

We contrast measurements of composite optical and ultraviolet (UV)
spectra constructed from samples of QSOs defined by their soft X-ray
brightness.  X-ray bright (XB) composites show stronger emission lines
in general, but particularly from the narrow line region. The
difference in the \oiii/\hb\, ratio is particularly striking, and even
more so when blended \feii\, emission is properly subtracted.  
The correlation of this ratio with X-ray brightness were
principal components of QSO spectral diversity found by Boroson \&
Green.  We find here that other, much weaker narrow optical
forbidden lines (\oii\, and \nev) are enhanced by factors of 2 to
3 in our XB composites, and that narrow line emission is  also strongly
enhanced in the XB UV composite.  Broad permitted line fluxes are
slightly larger for all XB spectra, but the narrow/broad line ratio
stays similar or increases strongly with X-ray brightness for all
strong permitted lines except \hb. 

Spectral differences between samples divided by their
relative X-ray brightness (as measured by \aox) exceed those seen
between complementary samples divided by luminosity or radio loudness.
We propose that the Baldwin effect may be a secondary
correlation to the primary relationship between \aox\, and emission
line equivalent width.  We conclude that either 1) \ew\, depends
primarily on the {\em shape} of the ionizing continuum, as crudely
characterized here by \aox\, or 2) both \ew\, and \aox\, are related
to some third parameter characterizing the QSO physics.  One such 
possibility is intrinsic warm absorption;  a soft X-ray absorber situated
between the broad and narrow line regions can successfully account for
many of the properties observed. 

\end{abstract}



\section{INTRODUCTION}
\label{intro}

Most QSOs have been discovered by virtue of their strong optical/UV
emission lines, or non-stellar colors in this bandpass.  Our
understanding to date of the violent inner regions of active galactic
nuclei (AGN) also derive in large part from their optical/ultraviolet
(OUV) spectra.
The production of QSO emission lines is widely attributed to
photoionization and heating of the emitting gas by the UV to X-ray
continuum (e.g., Ferland \& Shields 1985; Krolik \& Kallman 1988).
Studies investigating the relationship of emission lines to continuum
radiation have a long history in the field, but several strong
observational relationships remain unexplained.

If the proportionality between line and continuum strength were
linear, then diagnostics such as line ratios and \ew\, would be
independent of continuum luminosity.  Baldwin (1977) first noticed
that in high redshift quasars, the \ew\, of the CIV $\lambda1550$\AA\,
emission line in quasars decreases with increasing UV ($1450$\AA)
luminosity.  The Baldwin effect (BEff, hereafter) was also found to be
strong for ions such as OVI, NV, He\,II, CIII], Mg\,II, and Ly$\alpha$
(e.g., Tytler \& Fan 1992, Zamorani et al. 1992).  The initial excitement 
about the potential for the BEff as a standard candle and cosmological
probe has faded; the dispersion in the relationship is too large.
Neither the source of that dispersion, nor cause of the BEff itself
have been definitively identified. However, some possible explanations
for the BEff have been offered, one being a dependence of the
continuum spectral energy distribution (SED) on luminosity (Zheng \&
Malkan 1993, Green 1996).    

Many important lines respond primarily to the extreme ultraviolet
(EUV) or soft X-ray continuum.  Unfortunately, the EUV band is
severely obscured by Galactic absorption.  However, constraints on the
ionizing continuum are available through analysis of the adjacent UV
and soft X-ray windows.  In a small, uniform sample of
optically-selected QSOs (Laor et al. 1997), the strongest correlation
found between X-ray continuum and optical emission line parameters was
of the soft X-ray spectral slope
\ax\, (where $\fnu \propto \nu^{-\ax}$) and the FWHM of the \hb\, 
emission line.  Strong correlations between \ax, $L_{\oiii}$,
\feii/\hb, and the \oiii/\hb\, ratio were seen both there and in
previous QSO studies (e.g., Boroson \& Green 1992).  The latter
authors found that most of the variation in the observed properties of
low-redshift QSOs can be represented in a principal component analysis
by eigenvectors linking \feii, \oiii, \hb, and
\heii\, emission line properties and continuum properties
such as radio loudness, and the relative strength of X-ray emission,
as characterized by \aox\, (defined below).  A recent, and possibly
related result is that Seyfert 1s with broad
\hb\, emission lines tend to have hard (flat) X-ray spectral slopes
(e.g., Brandt, Mathur, \& Elvis 1997).  In a hard-X-ray-selected
sample of (mostly Seyfert) AGN, narrow \oiii\, flux correlates well
with X-ray flux, while broad Balmer lines do not (Grossan 1992).  The
physical origin of these diverse and interrelated correlations has
yet to be determined.  

We are launching a large-scale effort to probe these effects in
large samples, using both data and analysis as homogeneous as
possible.  Many physically informative trends intrinsic to QSOs may be
masked by dispersion in the data due to either low signal-to-noise or
variability.  An important tool for studying global properties of QSOs  
is the co-addition of data for samples of QSOs.  In this paper, we
concentrate on an analysis of composite optical/UV spectra of
subsamples of QSOs grouped by the relative strength of their
soft X-ray emission.

\section{Analysis}
\label{analysis}

\subsection{Constructing Comparison Samples}
\label{samples}

Although the signal-to-noise ratio (S/N) for the individual
spectra in both samples we study here tends to be about 10 or less per
resolution element, the co-addition (averaging) of spectra with
similar continuum properties allow us to increase the S/N and
constrain the properties of the average QSO.  Through averaging, a
greater number of emission lines, with a wider range of ionization
energies, and finer details in emission line profiles become measurable.  
This technique has been applied in several recent studies (e.g.,
Cristiani \& Vio 1990, Francis et al. 1991, Osmer, Porter, \& Green
1994, Zheng et al. 1996).  What is lost is a reliable measure of the
intrinsic dispersion in the observed correlations.  However,
it is important to first discover the correlations intrinsic to the
average QSO.  The sources of dispersion in the relationship can later
be studied if data of adequate S/N are available for a large enough sample. 

Table~1 summarizes mean continuum properties 
for the X-ray bright and X-ray faint subsamples we culled
from the LBQS and \IUE\, QSO samples described below.
Optical and X-ray luminosities are taken from Green et al. (1995) and
Green (1996), and assume $H_0=50$ km s\mone 
~Mpc\mone ~and $q_0=0.5$.  The slope of a hypothetical powerlaw
connecting  2500\,\AA~ and 2~keV is defined as $\aox\, = 0.384~{\rm
log} (\frac{ \lopt}{\lx })$, so that \aox\, is larger for objects with
stronger optical emission relative to X-ray.

\subsubsection{LBQS Sample}

The Large Bright Quasar Survey (LBQS; Hewett, Foltz \& Chaffee 1995)
is a sample of more than a thousand QSOs, uniformly-selected
over a wide range of redshifts.  LBQS QSO candidates were selected
using the Automatic Plate Measuring Machine (Irwin \& Trimble 1984) to
scan UK Schmidt direct photographic and objective prism plates.  A
combination of quantifiable selection techniques were used, including
color-selection, selection of objects with strong emission lines,
selection of objects having redshifted absorption features or
continuum breaks.  The technique appears to be highly efficient at
finding QSOs with $0.2<z<3.3$, a significantly broader range than past
work.  Follow-up ($6~-~10$\AA\, resolution) optical spectra with
S/N$\approx$10 (in the continuum at $\sim$4500\AA) were obtained at
the MMT and 2.5m duPont telescopes.  The digital spectra used here
were graciously provided to the author by Craig Foltz.

  Soft X-ray data for LBQS QSOs was selected from the \ros\, All-Sky
Survey (RASS) as detailed in Green et al. (1995).  Of the 908 QSOs in 
the LBQS/RASS sample, 92 are detected in X-rays. For the
non-detections, we assign an upper limit of 4$\sigma$ to the
raw counts.

Before generating composite spectra, we remove QSOs with known
broad absorption lines (BAL QSOs) from the LBQS/RASS sample. 
Above $z=1$, an insufficient fraction of QSOs are detected 
to construct X-ray bright and faint subsamples of comparable size.
We therefore exclude QSOs with redshifts greater than one,
for which the RASS data are insufficiently sensitive to be
of use.  To create composite spectra of similar S/N,
we choose a dividing point of \aox\, that 
results in similarly-sized subsamples.  We construct the X-ray bright
sample (XB hereafter, 60 QSOs) of detected objects
only, with $\aox < 1.475$.  The X-ray faint sample (XF hereafter, 
54 QSOs) includes both detections and lower limits
with $\aox \ge 1.475$.  QSOs with lower limits below that value could 
rightly belong either to the XB or XF sample, and so are excluded
from consideration.  Note also that the value of 1.475 does not
imply that \aox\, is measured to such accuracy.  Rather, it provides
a convenient dividing line near the median of the small range of
\aox\, values ($\sim 1.2$ to $1.6$) typically measured in QSOs.

Continuum properties of the final XF and XB
(\aox -selected) samples are listed in Table~1.  
In survival analysis, if the lowest (highest) point
in the data set is an upper (lower) limit, the mean is not well
defined, since the distribution is not normalizable, and so the
outlying censored point is redefined as a detection. For the XF sample,
we list here the value of that redefined limit, where the distribution
is truncated.  The resulting mean value is biased, so that the true
mean values of \aox\, and \loglx\, for the LBQS XF sample are probably
even more X-ray faint than those listed. 

We find no significant differences in the distributions
of redshift, \nhgal, or \lopt\, between our XB and XF subsamples
of the LBQS. In any case, the emission line properties of the LBQS as
a whole show no strong dependence on either luminosity or redshift
(Francis et al. 1993). 

\subsubsection{\IUE\, Sample}

To explore changes in QSO UV spectra as a function of \aox,
we include a previously compiled sample of QSOs observed by both the
{\em International Ultraviolet Explorer} (\IUE) and \ein.
This sample was selected as described in Green (1996), by requiring
that the QSOs in the \IUE\, sample have \ein\, soft X-ray data
available in Wilkes et al. (1994), we define the \IUE/\ein~ sample of
49 objects.    

Again, we remove all known BAL QSOs from the \IUE\, sample.
With this sample, due to selection effects, it is not possible
to achieve similar-sized subsamples with similar redshift
distributions by selecting an appropriate \aox\, value at which
to divide the sample.  Similar redshift distributions are obtained
for $\aox=1.3$, but then the XB subsample contains only about a dozen
QSOs, compared to 37 in the XF subsample.  Similar subsample
sizes are obtained for $\aox=1.4$, for which, however, the
requirement of a detection in the XB sample results
in a lower mean redshift, due mostly to extra QSOs between
$0.1<z<0.2$.  Since QSO spectral
evolution between their mean redshifts (0.34 and 0.57, respectively)
is negligible, we emphasize the subsamples split at  $\aox=1.4$,
which is in any case closer to the dividing value for the LBQS
subsamples.  However, we also check the subsamples split at
$\aox=1.3$, to insure that evolution or luminosity effects
do not significantly bias our results.

\subsection{Constructing Composite Spectra}
\label{composites}

We first deredshift each individual spectrum by dividing the linear 
dispersion coefficients (initial wavelength and wavelength per pixel)
by $(1+z)$.  An estimate of the restframe continuum flux $f_{c,l}$ at
a chosen normalization point is then derived, by fitting a linear
continuum between two continuum bands. The width and center of these
bands, typically chosen to straddle an important emission line,
are listed for each wavelength region in Table~2.  We
then normalize the entire spectrum via division by $f_{c,l}$.

Each normalized spectrum is first rebinned to the dispersion of the
composite spectrum, conserving flux via interpolation.
We choose 2.5\AA\, bins, similar to the majority of the 
individual LBQS spectra.  The \IUE\, spectra are binned to 1.18\AA, as 
in the original short wavelength prime (SWP) spectra.  Each
normalized, rebinned spectrum 
is then stored as a vector in a 2-D array.  Finally, for each pixel in 
the completed array, the number of spectra $N$ with fluxes in
that restframe wavelength bin is tallied.  If $N\geq 3$, the median
and the mean of all the flux values in the bin are computed
and stored in the final 1-D composite spectra, as is $N$ in the
final histogram array.  A similar procedure is performed for each bin,
until the spectra are completed.  The strong sky line at $\lambda
5577$\AA, sometimes poorly subtracted, was omitted from the LBQS
composite. Geocoronal \lya\, lines were similarly excluded when
building the \IUE\, composites.

We chose four separate continuum points to generate LBQS composite
spectra in four wavelength regions (Table~2).
Composites normalized at different continuum points may have
different histograms, since they require the restframe continuum bands
to be present in all contributing observed frame spectra.  Histograms
of the number of QSOs contributing to each LBQS composite are shown
in Figure~1.  The \IUE\, composites were all
normalized at the wavelength of \civ, and the histograms for
those spectra are shown in Figure~2.

\begin{figure*}[!h]
    \leavevmode
\epsfbox[52 144 392 718]{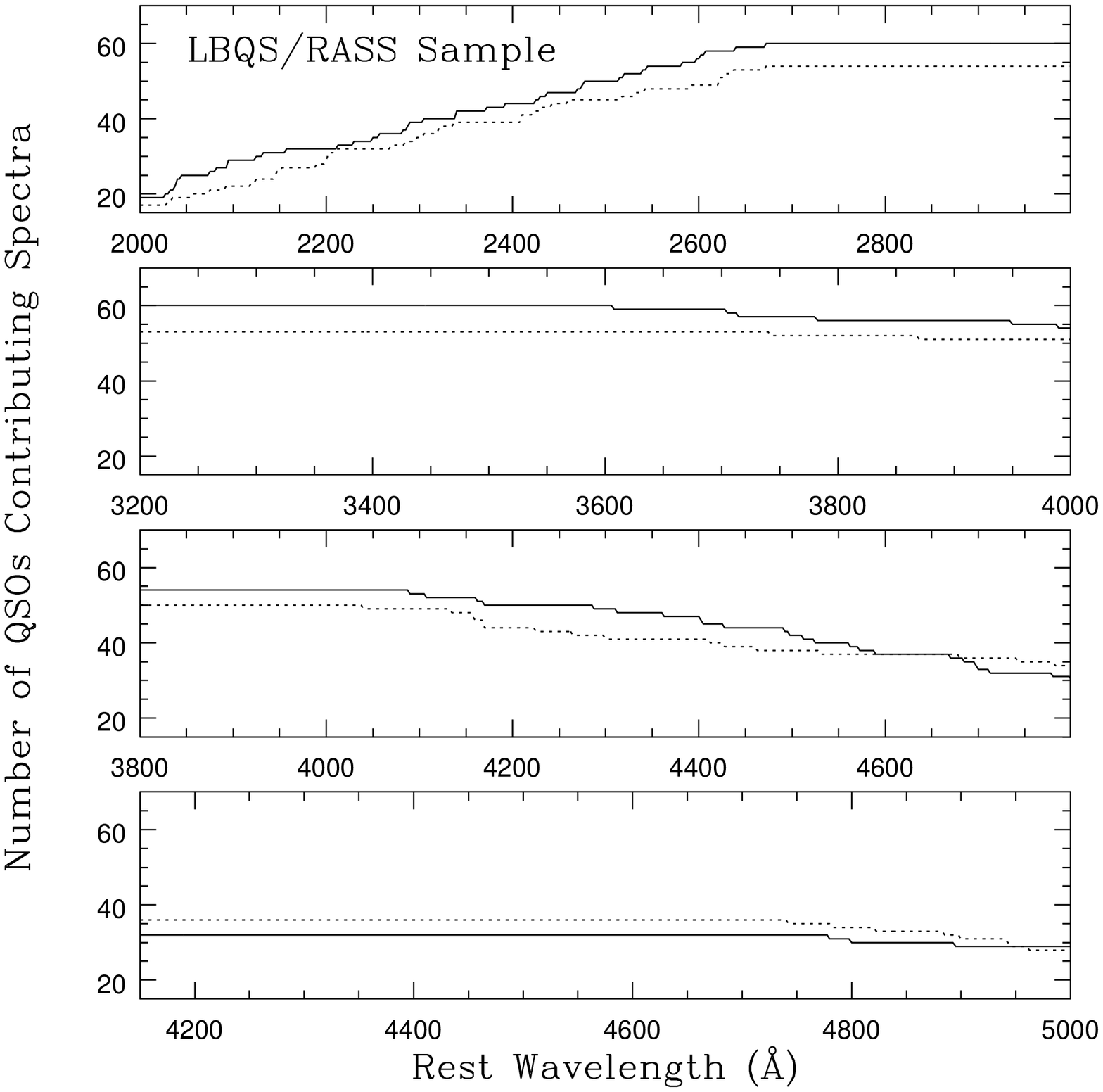}
\figcaption[]{Histograms of the number of QSOs whose
spectra contribute to each of 8 LBQS composite spectra.  
Four separate continuum points were used to generate LBQS composite
spectra for XB and XF subsamples in 4 separate wavelength regions
(see Table~2). Composites normalized at the four different continuum
points have different histograms, since they require the restframe
continuum bands to be present in all contributing observed frame
spectra.  Different selection criteria for the XB (solid line), and
the XF subsamples (dashed line) also yield slightly different
histograms.
\label{flbqshistos}}
\end{figure*}

\begin{figure*}[!h]
    \leavevmode
\epsfbox[52 144 392 718]{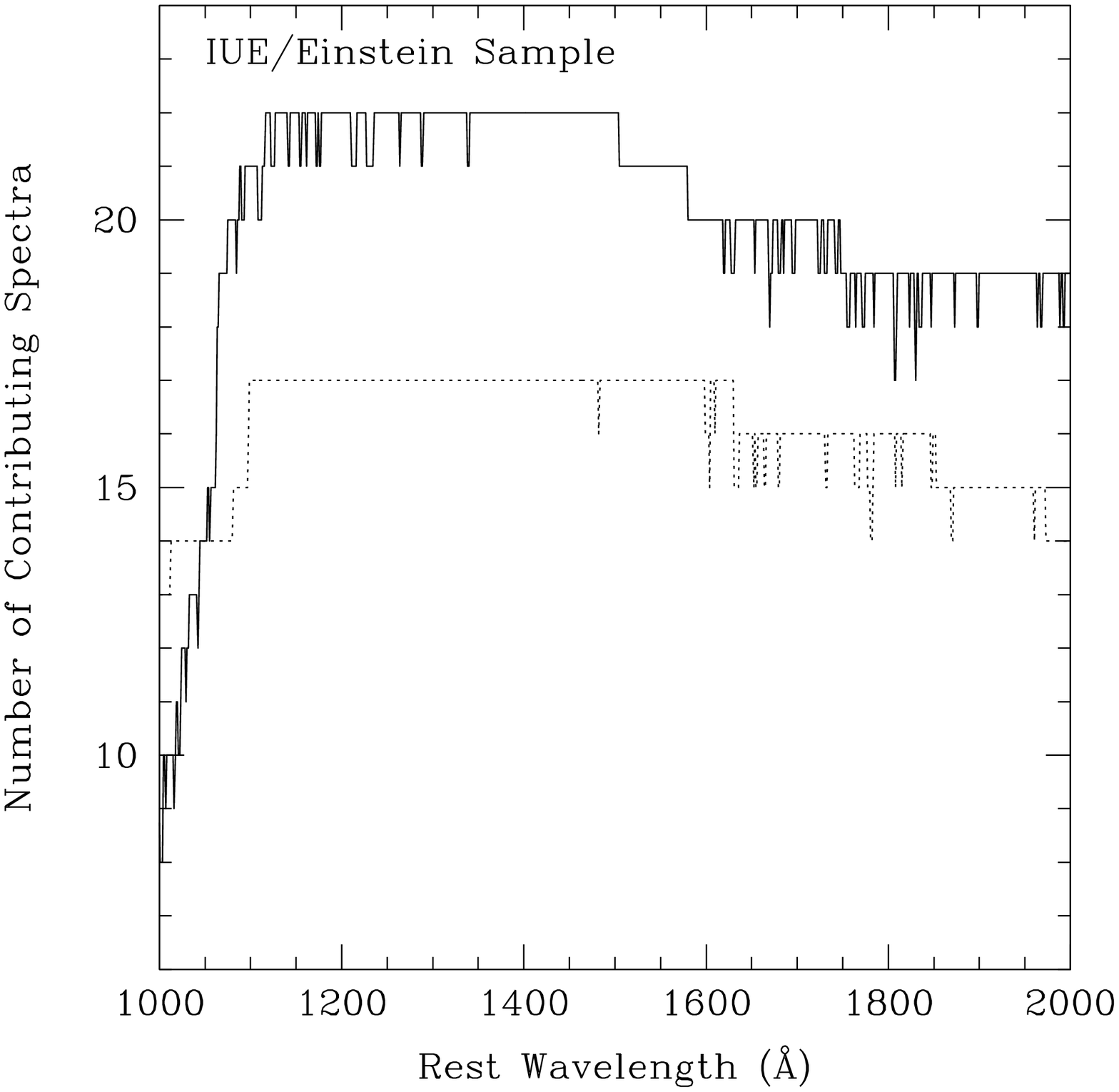}
\figcaption[figs/iuehistos.ps]{
Histograms of the number of QSOs whose
spectra contribute to the \IUE\, composite spectra for the XB 
subsample (solid line), and the XF subsample (dashed line). The
\IUE\, composites were all normalized at the rest wavelength of \civ.
\label{fiuehistos}}
\end{figure*}

The median composites appear to be smoother than the means, since they
are less affected by spikes and low S/N features in the individual
spectra. As a result, the $\chi^2$ values of model fits are also lower
for median composites.  However, since the median omits QSOs with more
extreme spectral properties, we prefer to analyze the average composites.
We check the significance of every strong difference between XB and XF
average composites by measurement of the median composites, to
determine whether outliers dominate the feature in question.

\section{Emission Line Measurements: LBQS}
\label{optlines}

No analysis of continuum slope is presented for the LBQS spectra,
since these observations are not fully spectrophotometric.  
Using the IRAF task SPECFIT (Kriss 1994), we fit simple empirical
models to the LBQS composite spectra in four wavelength regions,
defined in Table~2.  
In each region, we begin by fitting a powerlaw (hereafter, PL)
continuum using only comparatively line-free regions (listed
for each region below).  The resulting
continuum fit parameters are slope $\alpha$ and intercept $f_{1000}$,
such that flux $\flam = f_{1000}(\frac{\lambda}{1000})^{\alpha}$.
Since the LBQS spectra are not spectrophotometric, the
continuum fit results should be used only to derive exact
composite equivalent widths from (normalized) line fluxes if desired.
The intercept value reflects the arbitrary normalization of individual
spectra to unity at the chosen wavelength $\lambda_n$
shown in Table~2.  

The emission line components are assumed to be Gaussian and symmetric
(skew fixed at unity), so output from the fits includes flux,
centroid, and FWHM for each line.  Results from these fits are shown
in Table~3 and Figure~3.  The errors listed in both Table~3 and 4
are directly from the SPECFIT task, which assumes that the errors on
the input spectrum follow a Gaussian distribution. The tabulated
errors represent one sigma for a single interesting parameter.

\begin{figure*}[!h]
    \leavevmode
\epsfbox[52 144 392 718]{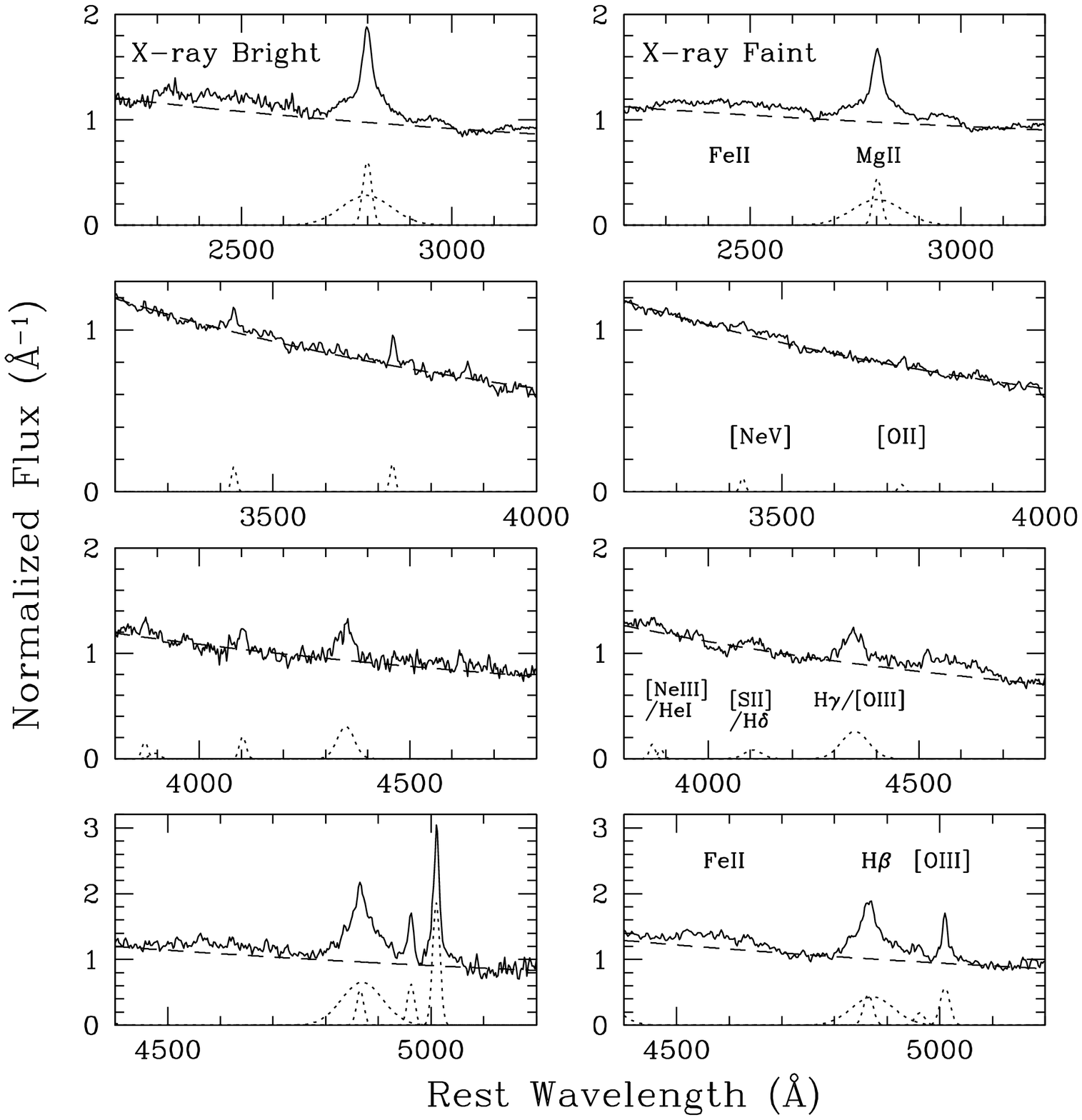}
\figcaption[figs/opt2.ps]{
Final normalized composite spectra of the LBQS sample.  
The left column contains plots of the XB composites,  
while the right column shows XF composites. 
Each row shows a portion of a composite constructed from spectra
normalized at a different rest wavelength continuum point (see
Table~2):  
$\lambda$2798 (top row),
$\lambda$3426 (second row),
$\lambda$4102 (third row),
$\lambda$4863 (bottom row).
Individual lines and their parameters measured using SPECFIT (Kriss 1994) 
are listed in Table~3.  Fitted components are shown in
each plot.  These include the best powerlaw continuum fit to
relatively line-free continuum regions (dashed line) and Gaussian
emission lines (dotted lines). Lines are identified by name in the
X-ray faint plots. 
\label{foptfit}}
\end{figure*}

\subsection{The \hb\, and \oiii\, Region}
\label{hbregion}

We fit a PL continuum using these relatively 
line-free regions: $\lambda$4020-4050, 4150-4270, 4420-4450, 4710-4780,  
and 5070-5130.  The resulting fit parameters are fixed,
while the following single Gaussian components are fit to the
strong emission lines between 4150 to 5100\AA:
1) Broad \hb;
2) Narrow \hb, with FWHM fixed to that of \oiii$\,\lambda$5007;
3) \oiii$\,\lambda$5007;
4) \oiii$\,\lambda$4959, fixed relative to the central wavelength and
(1/3) flux of component (3);
Finally, the blended iron emission above the PL continuum
is summed over the wavelength range $\lambda\lambda 4434-4684$.

As can be seen from Figure~3 and Table~3,
by far the strongest effect is the great 
relative strength of \oiii\, emission in the XB composite; \oiii\,
emission is $\sim 2.5$ times stronger in X-ray bright sample.
Broad \hb\, emission is about 40\% stronger in the XB composite,
although the FWHM are similar at $\sim$5700\kms.  Strengths of the
narrow component of \hb\, are comparable.  In both XB and XF
composites, broad \hb\, is redshifted relative to rest (\i.e.,
relative to \oiii$\,\lambda$5007 or narrow \hb) by about 300\kms.

Boroson \& Green (1992; hereafter BG92), in a principal
component analysis of low redshift QSO spectra, found that the
principal eigenvector of their sample (the linear combination
of parameters that represents the largest variance in the sample)
is primarily an anticorrelation between the strength of \oiii\, emission
and optical \feii\, strength.   This trend is in the
same direction as we find here, with the large \woiii, XB sample
having the lower \feii\, equivalent width, as judged by
the iron emission near $\lambda 4600$\AA\, (Table~3).   


We can make more precise measurements in this region if we subtract
\feii\, emission.  Marianne Vestergaard has accomplished this task,
using a similar technique to BG92.  A narrow optical \feii\, template
spectrum (of IZw1, from BG92) is convolved with a Gaussian function and
subtracted from the LBQS composites, fitting iteratively by eye.  The
flux in the iron template between $\lambda 4434$ and $\lambda 4684$ is
within $\sim 10\%$ of that found via our simple summation technique.
Due to a strong iron multiplet with lines at
$\lambda\lambda$4924,5015, and 5169, however, subtraction of \feii\,
decreases the apparent \woiii\, in both the XF and XB composites.  In the XF
composite, where iron emission is strong, the measured \woiii\, at
$\lambda 5007$ decreases from 13.5 to 7.5, and its measured FWHM from
1330 to 650\kms.  The XF \hb\, line changes by comparison only
slightly after iron subtraction (from a total equivalent width of 53
to 47).  In the XB composite, where iron emission is weak, the
measured change in equivalent width of both
\oiii\, and \hb\, after \feii\, subtraction are within the errors
($\Delta\ew <3$\AA), as are the changes in FWHM.  Since the strength
of \oiii\, in the XF composite is initially low, {\em the difference in
the measured \oiii/\hb\, ratio between XF and XB composites is
substantially increased by \feii\, subtraction}.  Much smaller effects
are expected on other emission lines treated here.  An important
extension of this technique to the UV will be described in an upcoming
paper (Vestergaard \& Green 1998).

\subsection{The \hd\, Region}
\label{hdregion}

  First, we fit a PL continuum through the windows:
$\lambda\lambda$3790-3845,  4020-4050,  4150-4270,  4420-4450, and
4710-4780.  Gaussian components are fit between
3800 - 4800\AA\, to the following lines:  
1) \neiii$\,\lambda$3869;
2) \hei$\,\lambda$3889; 
3) \sii/\hd$\,\lambda$4068/4076+4102;
4) \hg/\oiii$\,\lambda$4340+4363
The only significant difference seen between emission lines
in this region is that the \hg/\oiii\, blend is {\em stronger}
(at the 2.7$\sigma$ level) in the XF composite.

\subsection{The \nev\, and \oii\, Region}
\label{neregion}

Line-free regions at $\lambda$3200-3400, 3500-3700, and 3800-4000
are used for the PL fit.  Then single Gaussian components are fit for
1) \nev$\,\lambda3426$;
2) \oii$\,\lambda3727$. 
As can be seen from Figure~3 and Table~3,
both the \nev\, and \oii\, line fluxes are much
stronger in the XB composite, consistent with the relative strengths
of other unblended narrow lines measured. For the XF sample, the very
weak \oii\, line could not be successfully 
fit when leaving both $\lambda_c$ and FWHM free to vary, so we fixed 
the central wavelength at 3727\AA.  Similarly, the weakness of 
\nev\, demanded that its FWHM be fixed, because otherwise
SPECFIT seeks to fit neighboring features with an unrealistically broad line.
The use of composite spectra enables us here for the first time
to show that these narrow lines are about 3 times as strong
in the X-ray bright QSOs as in their X-ray faint analogs.

\subsection{The \mgii\, Region}
\label{mgregion}

Here a PL continuum is fit to $\lambda$2195-2275, 2645-2700, and
3020-3100 windows. The spectrum from 2000 and 3200\AA\, is then
fit with the fixed PL plus these Gaussian components:
1) narrow \mgii;
2) broad \mgii, fixed in wavelength to the narrow component.
Without fixing the broad line width, the fit would wander
off trying to include nearby blended iron lines.
Finally, the blended iron emission above the PL continuum
is summed over the wavelength range 2210-2730\AA.

We find here that the XB sample has stronger UV  \feii\, emission
(Table~4), opposite to the trend for optical \feii.
Green et al. (1995) also found that QSOs in the LBQS with strong
UV Fe\,II emission are particularly bright in the soft X-ray
bandpass.  UV \feii\, lines have their principal ionizing
and heating continuum above 500eV; Fe\,II lines
in the optical are principally due to continuum above 800eV.

\section{Emission Line Measurements: \IUE}
\label{uvlines}


For the \IUE\, spectra, convincing model fits to the composite
spectra could be obtained when fitting a simple PL continuum simultaneously
with lines.  The following single Gaussian components are fit to the
spectrum between 1000 to 2000\AA : 
1) \lyb/\ovi;
2) broad \lya;
3) narrow \lya;
4) NV;
5) OI;
6) \siiv;
7) broad \civ;
8) narrow \civ;
9) \civ\, absorption.
Again, Gaussian components are assumed to be symmetric.
The data do not warrant multiple Gaussian fits to lines as weak or
weaker than NV.  However, some such lines are included in the fit
partly to insure that a reasonable continuum estimate is derived.  

Results from these fits, shown in Table~4 and Figure~4, are described below. 

\begin{figure*}[!h]
    \leavevmode
\epsfbox[52 144 392 718]{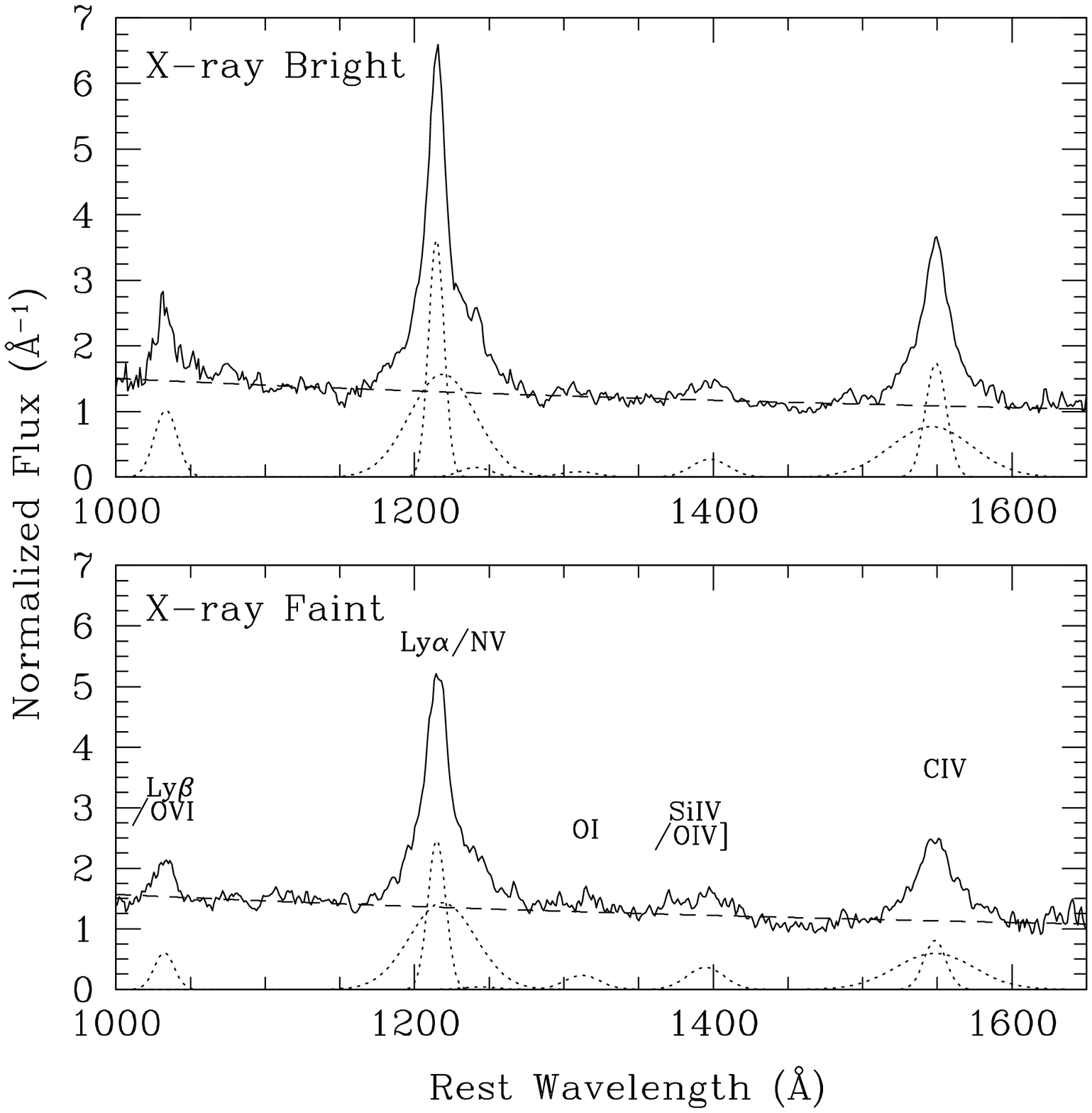}
\figcaption[figs/uv.ps]{
Final normalized composite spectra of the \IUE/\ein\, sample.  The top
plot is the XB composite, and the bottom shows the XF composite.  Both
these \IUE\, composites were constructed from spectra normalized at
$\lambda 1549$, the rest wavelength of \civ.  Individual lines and
their measured parameters using SPECFIT (Kriss 1994) are listed in Table~4.
The best-fit components are shown in each plot.  These include a
powerlaw continuum (dashed line), fit simultaneously with
Gaussian emission lines (dotted lines). Lines are identified by name
in the X-ray faint plot.  Note the apparent continuum deficit centered
near $\lambda 1480$\AA\, in the XF composite.
\label{fuvfit}}
\end{figure*}

1)  The PL continuum fits are similar in slope.  At the
2$\sigma$ level, the XF sample may be marginally steeper.  If
verified in other samples, this could be caused by reddening or
absorption by dust.  However, the similarity of the slopes
suggests that the relative attenuation between the XB and XF
spectra is grey.  

2) Just as for the LBQS composites, we find for the \IUE\, sample
that line emission is significantly stronger in the XB composite.
\lyb/\ovi, \lya, and \civ, when fit with a single Gaussian component, are 
each significantly ($>3\sigma$) stronger in the XB composite.
For \lya, the narrow/broad line flux ratios are higher in the XB
composite as well.  Some marginal evidence ($\sim 1.5\sigma$)
for the same effect is seen for \civ.   

From Table~4 it is clear that the \lyb/\ovi\, emission line shows a
large contrast. The inverse correlation between \lyb/\ovi\, and
\aox\, was also remarked recently in a sample of individual HST FOS
spectra (Zheng, Kriss \& Davidsen 1995).  Among weaker emission lines,
OI$\,\lambda 1302$ and SiIV/OIV] show a marginal ($<2\sigma$) trend in
the opposite direction, with smaller \ew\, in the XB composite.
There is also a suggestion that in the XF composite, NV is weak or
nonexistent.  These should be investigated at higher S/N.

3) There is marginal evidence for a continuum flux deficit blueward of
\civ\, of the XF continuum.  

\subsection{Evidence for Absorption}
\label{abs}

  It proved difficult to fit the \IUE\, XF spectrum with a convincing
PL continuum, because of a dip blueward of the \civ\, emission line,
which we tentatively interpret as absorption.  
We fit a simple Gaussian component to the absorption, which is sufficient to
characterize the absorbed flux and velocity
width. The result is that the CIV emission line fits remain unchanged,
while the overall fit improves, but only for the XF spectrum.
(The formal improvement in $\chi^2$
would appear marginal because the absorbed region covers only a few
percent of the spectrum.)  The equivalent width (\ew) of the
absorption is $8.8\pm1.4$\AA, centered at  $1480\pm10$\AA, with FWHM of
$12500\pm2800$km\mone\, (see Table~4).  Some evidence for a
similar dip can be seen even in the XB spectrum, but it is 
smaller, and its \ew\, is not significant in the fit in comparison to the
formal errors ($<1.5\sigma$).   

If individual absorbers contribute to the dip in the \IUE\, XF
composite, they are likely to be narrow, weak, and undetected in the noise
of the individual spectra.  When the spectra of XF QSOs are combined,
however, the enlarged width and higher S/N of the composite absorber
could enable detection of a broad feature. High velocity intrinsic narrow
line absorption in QSOs is just now being recognized in individual
QSOs (Hamann et al. 1997); the challenge is to properly distinguish it
from intervening absorption.

In summary, there is evidence for a dip blueward of the
\civ\, emission line in the XF composite, independent of any
reasonable continuum choice.  If the dip in the XF composite is due to
\civ\, absorption, the absorber is highly ionized, and probably
situated near the ionizing source.  The large velocity width also
suggests proximity to the broad line region (BLR). The weak X-ray
emission combined with evidence for high velocity ionized absorbers is
reminiscent of recent results associating soft X-ray and UV absorption
(Mathur 1994, Mathur, Elvis \& Wilkes 1995, Green \& Mathur 1996,
Green 1997).  

\section{Discussion}
\label{discuss}

We now consider a variety of possibilities to account for the
optical and UV spectral difference between X-ray bright and X-ray
faint QSOs; 
1) luminosity effects,
2) radio loudness,
3) absorption, and
4) changes in the {\em intrinsic} spectral energy distribution (SED).
The strength of some of these effects is directly testable
using the samples at hand.  

\subsection{Luminosity Subsamples}
\label{lum}

Both selection effects and secondary correlations must be considered
when evaluating the significance of observed correlations such as
these.  Two well-known effects could conspire to produce an
overall weakening of emission lines with increasing \aox.
First, \aox\, is known to increase with optical luminosity
(Wilkes et al. 1994, Green et al. 1995), at least for
optically-selected samples (LaFranca 1995).  Secondly, as luminosity  
increases, line equivalent width decreases (i.e., the Baldwin Effect;
Baldwin 1977).  Could these effects combine to produce the
anti-correlation of \aox\, and line strength observed here? 

As can be seen from Table~1, the LBQS subsamples are
well-matched in optical luminosity, so that no Baldwin effect
is expected.  Furthermore, the strength of the Baldwin effect in the
optical is known to be weak.

Our results from the \IUE\, subsamples are less immune to 
a Baldwin effect/SED conspiracy for the following reasons:
1) our \IUE\, subsamples are not as well-matched in luminosity, and
2) the Baldwin effect is much stronger in the UV.
We therefore perform a stringent check, by applying the same
spectral averaging techniques now to new subsamples defined by UV
luminosity.   

We divided the \IUE\, sample at the mean UV luminosity
\footnote{The log rest-frame luminosity \logluv, in \fnucgs\, at
1450\AA, is defined in Green (1996)} value of 30.6.  The 
resulting low UV luminosity (UVLO; 27 QSO) and 
high UV luminosity (UVHI; 22 QSO) subsamples both had mean
$\aox=1.4\pm .04$.  SPECFIT procedures identical to those of the XB
and XF samples were applied. Virtually all spectral differences were
{\em less} significant in the UV luminosity subsamples than in the \aox\,
subsamples.  Only narrow \lya\, emission changes more strongly
between UV subsamples than between the \aox\, subsamples.  Indeed, the
Baldwin Effect appears to be strongest in the narrow line components
of both \lya\, and \civ.  The bulk of the effect could be 
due to differences in narrow line region (NLR) emission, as also suggested
by Osmer, Porter, \& Green (1994).

Since emission line correlations are stronger with \aox\, than with
luminosity, we conclude that either 1) \ew\, depends primarily on the
{\em shape} of the ionizing continuum, crudely characterized
here by \aox\, or 2) both \ew\, and \aox\, are related to some third
parameter characterizing the QSO physics.  One such possibility is
absorption.

Could the correlation of \aox\, to luminosity {\em cause} the
Baldwin effect?  Although by design we have selected subsamples of
similar luminosity for our XF and XB composites, 
we may suppose that the primary relationship between \ew\,
and \aox\, is propagated into the relationship between \aox\,
and luminosity, and test the strength of the predicted secondary
relationship between \ew\, and luminosity that results, i.e. the
Baldwin Effect.  

We begin by simply contrasting the observed response here of
\wciv\, and \wmgii\, to \aox\, with that predicted in the most recent
comprehensive study of the Baldwin effect in optically-selected QSOs
(Zamorani et al. 1992).  The change $\Delta\overline{\aox}\sim 0.3$
between XF and XB subsamples is similar in our \IUE\, and LBQS samples.  
The change in log\,\wmgii\, predicted by
the BEff in this line is $\Delta{\rm log}~\wmgii) \sim -0.68\aox =
0.18$.  The change we actually 
measure between XB and XF composites is (from
the values in Table~3) $\Delta{\rm log}~\wmgii=1.0$.
Similarly, the change predicted by the BEff in log\,\wciv\, seen by
Zamorani et al. is $\Delta{\rm log}~\wciv \sim -1.16\aox = 0.30$,
while we actually measure $\Delta{\rm log}~\wciv=1.57$. Thus, the
effect of \aox\, on emission line strengths is some 5 -- 6 times
larger than that expected if it were secondary to a luminosity effect.
Indeed, {\em it seems likely that the BEff is secondary to the
relationship between \aox\, and line equivalent width.}

If the BEff is caused by a change in \aox\, concomitant with
luminosity, then the strongest \ew(\aox) relationship we see,
that of $\woiii$, predicts a BEff in \oiii.  This is indeed seen
in the BG92 data, where the probability of no correlation
(the null hypothesis) between $M_V$ and \woiii\,$\lambda 5007$ is
$<1\%$.

\subsection{Radio Properties}
\label{radio}


Radio and X-ray loudness are correlated;  Green et al. (1995)
confirmed that RL QSOs are more soft X-ray luminous than RQ QSOs
in the LBQS.  Unfortunately, the difference in radio loudness between the
LBQS XF and XB subsamples cannot be well-characterized: although 4
(of 17 with radio data) are radio-detected in the LBQS XB subsample,
only one QSO (of 38) in the XF subsample is a radio detection.  

However, OUV spectral differences as a function of radio loudness have been
extensively studied.  Differences between emission lines in radio loud
QSO (RLQ) and radio quiet QSO (RQQ) spectra longward of 1600\AA\, are
quite subtle, and there is a 
remarkable similarity in \mgii\, and \ciii\, emission lines
between RLQs and RQQs (Francis et al. 1993).  Distinctions found
by BG92 include a redward asymmetry of \hb\, in RLQs, 
whereas RQQs show about equal numbers of red and blue asymmetries
(BG92). RL QSOs tend to have strong \oiii\, and weak optical \feii.
In Wang, Brinkmann, \& Bergeron (1996), neither \hb\, nor 
optical \feii\, measurements correlate with radio loudness.  

To check the effects of radio loudness on UV emission lines, we
created RL and RQ composite UV spectra from the \IUE/\ein\, sample. 
We determined radio loudness from Falcke et al. (1996) for all
the UV excess selected (PG) QSOs, and from Veron-Cetty \& Veron (1993)
or the NED database for others.  Two QSOs are omitted because of intermediate
radio loudness, and two more have no published radio data.
Radio subsamples are identical in luminosity and well-matched in
redshift if we impose $z>0.15$.  This yields N=26 and
$\overline{z}=0.48$ for the RL subsample, N=13 and
$\overline{z}=0.52$ for RQ QSOs, and \loglopt=30.9 for both. 
There is some remaining difference between \aox\, distributions for
the \IUE/\ein\, radio subsamples, as expected, such that the RL
sample is somewhat more X-ray bright ($\overline{\aox}=1.36\pm0.04$)
than the RQ sample ($\overline{\aox}=1.47\pm0.05$).  

We find \lya\,emission to be somewhat stronger in the RQ composite.
\civ\, emission is of similar strength and FWHM in both RQ and RL
composites, but displaced slightly to the red in the RL composite.
Similar results were found independently by Wills \& Brotherton
(1995).  Thus overall, the emission line trends due to radio loudness
in our sample are weaker than, and tend to {\em diminish} those seen
in \aox.  The important result is that {\em spectral differences
between composites binned by \aox\, are significantly larger than
between composites binned by radio loudness}.

\subsection{Absorption and the \aox(\lopt) Relation}

 Absorption by ionized gas near the nucleus can extinguish soft X-ray
emission without having significant effects on the observed optical
emission. The BAL QSOs represent an extreme example, wherein observed
soft X-ray fluxes are at least an order of magnitude below those of
non-BAL QSOs of similar optical brightness (Green \& Mathur 1996).
Even AGN with much less spectacular UV absorption (e.g. narrow
absorption lines) show significant soft X-ray absorption (Mathur
1994, Mathur, Elvis, \& Wilkes 1995).   Since most QSOs have either
low S/N, low resolution, or no available UV spectra, many such absorbers
await recognition in current QSO samples.

Is it possible that the \aox(\lopt) correlation is itself caused by
absorption?  The hypothesis might be tested if {\em all} types of absorbed
QSOs were removed, including QSOs with known narrow-line intrinsic,
damped \lya, and \lya\, forest absorbers.  
A higher S/N sample that has been systematically searched for
absorption in both UV and soft X-ray spectra is needed.  We
are building such a sample from the HST FOS and ROSAT public
archives, and will study these issues in an upcoming paper.

We note that if at least some of the correlation is caused by 
increased warm absorption at higher luminosities, then the correlation
would be expected to be further weakened in soft X-ray selected
samples.  More common absorption in the UV bandpass
is suspected anecdotally at higher luminosities, but remains to be
demonstrated statistically.  In the X-ray bandpass, the same can be
said for the popular assumption that absorption is {\em less}
common at high luminosity.  This latter notion should be particularly
suspect, since higher energy rest-frame X-ray emission is observed in
most high luminosity objects, and requires a higher intrinsic absorbing
column for detection.  Furthermore, the most luminous objects in
samples to date have also been radio-loud (Reynolds 1997).

Variability may play a significant role in the 
\aox(\lopt) correlation.  Typically, optically-selected AGN are found
near the survey flux limit and therefore preferentially in a bright
phase, and then later followed up in an X-ray pointed observation.
Overall, more distant (and thus more luminous) optically-selected QSOs
would tend to have larger measured \aox. Even if variability-related biases
are not responsible for the observed \aox(\lopt) correlation,
such variability may be expected to introduce scatter into the
true relation.  As such, $\alpha_{ix}$ (the powerlaw spectral index between
1$\mu$m and 2~keV) might prove a more variability-resistant measure.
Indeed, Lawrence et al. (1997) have found that some of the primary
correlations discussed here become more significant when
$\alpha_{ix}$ replaces \aox.  Not only variability, but also
absorption more strongly affects the optical than the infrared.

\subsection{Intrinsic Spectral Energy Distributions}
\label{seds}

We have offered one interpretation of our measurements
that assumes that the {\em intrinsic} broadband continuum
emission from the QSO central engine is constant in shape,
independent of luminosity, but that the spectral
energy distribution (SED) seen by the NLR and/or by us 
may be strongly affected by intervening, possibly ionized absorbing
clouds.  However, the change in \aox, and the accompanying
changes in emission lines may at least in part be due to 
changes in the intrinsic SED.

 Many QSOs have soft ($\lapprox$1~keV) X-ray emission that exceeds the
extrapolation from the powerlaw continuum observed at higher energies
(e.g., Turner \& Pounds 1989, Masnou et al. 1992). This X-ray `soft
excess' has often been interpreted as the high energy continuation of
the UV/EUV/soft X-ray ``big blue bump'' (BBB), possibly thermal
emission from the surface of an accretion disk (although see Barvainis
1993).  Several workers (beginning with Malkan \& Sargent 1982)
have proposed as an explanation of the Baldwin effect that as QSO
luminosity increases, the BBB shifts toward lower energies at higher (OUV)
luminosities.  As QSOs become more luminous in the optical/UV band,
they thereby undergo a weaker increase in \lx, and therefore an increase in 
\aox.  The response of line flux and \ew\, depends in a fairly complicated
manner on the peak energy of the BBB, on the BBB normalization
relative to the powerlaw continuum, and on the ionization and heating
continuum of the line species in question.  Detailed photoionization
modeling using a variety of input continua, impinging on an
ensemble of clouds from the broad to the narrow line region,
including full self-shielding and optical depth effects are called for
(see e.g., Korista et al. 1997, Baldwin et al. 1995, Shields et
al. 1995).  On average, the strongest effect may be that higher
luminosity QSOs may undergo spectral evolution such that fewer photons
from a soft X-ray excess/BBB component are available for ionization. 

 A shift of the BBB to
lower energies at higher luminosity implies that the soft X-ray excess
should decrease to higher luminosities.  An apparent hardening of soft
X-ray PL spectral index has been seen in composite ROSAT spectra
of LBQS QSOs toward higher redshifts (e.g., Schartel et al. 1996) that
gibes with this picture, but again, higher energy rest-frame X-ray
emission is observed in most high luminosity objects, where a
higher intrinsic column is required before absorption can be detected.
In this picture, a stronger BEff might be expected for species of
higher ionization energy. There is some evidence for such a trend
(Zheng et al. 1992).  The intrinsic SED model does not predict
absorption features, but does imply that the \aox(\lopt) relation
should persist even in soft X-ray-selected samples.  

\section{Summary}
\label{summary}

By contrasting the composite optical/UV spectra of large samples of X-ray
bright and X-ray faint QSOs we have unveiled significant 
new correlations with X-ray brightness as characterized by \aox.
We find that \oiii\, emission in at least 2.5 times stronger
in our XB sample.  Proper subtraction of \feii\, suggests a true ratio
closer to 5.  We find that other, much weaker narrow optical forbidden lines
(\oii\, and \nev) are enhanced by factors of 2 to 3
in our composites. Narrow line emission is also strongly enhanced in
the XB UV composite.  Broad permitted line fluxes are slightly larger
for both XB composites, but velocity widths in the broad emission line
region are not significantly affected.  The narrow/broad line ratio
stays similar or increases with X-ray brightness for all strong lines
{\em except} \hb.   

We also find that UV \feii\, and optical \feii\, emission 
are correlated in opposite senses with \aox.  Optical \feii\,
equivalent widths decrease with X-ray brightness, while UV
iron equivalent widths increase.  This confirms similar suggestions 
elsewhere (e.g., Green et al. 1995, Lipari 1994, Boroson \& Meyers
1992).  Proper modeling of non-radiative heating, optical depth
effects, and iron recombination rates, along with improved
methods of \feii\, emission measurements may all be needed to
contribute to the solution of this intriguing observation.

Broad emission lines differ less between XB and XF QSOs
than do narrow lines.  Our interpretation of the data is that
absorbers tend to obscure the line of sight to the central ionizing
source both from us, and from the NLR.  A similar conclusion
indicating that less-altered continuum radiation reaches the BLR 
was recently suggested from different lines of evidence by Korista,
Ferland, \& Baldwin (1997).   

Our tests on complementary subsamples indicate that spectral
differences between subsamples divided by \aox\, exceed those seen
between samples divided by luminosity or radio loudness.  In
particular, we propose that the Baldwin effect may be a secondary
correlation to the primary relationship between \aox\, and emission
line equivalent width.  One test of this is that the Baldwin effect
should be dominated by narrow line components.

We note that for 23 UV-excess selected (PG) QSOs observed in the ROSAT
bandpass (Laor et al. 1997), correlations of emission line parameters
(FWHM(\hb), and optical \feii and \oiii\, strengths) are clearly
stronger with \ax\, than with \aox.  However, a correlation exists
between \aox\, and \ax; X-ray bright QSOs (with small \aox) tend to
have flat (hard) \ax.  At these and slightly lower luminosities
(Seyfert~1s, e.g., Boller et al. 1996), only objects with flat (hard)
\ax\, have large FWHM(\hb).  If \hb\, is representative, the
combination of these two trends would predict that XB QSOs should on
average have broader emission lines, counter to the overall trends
seen in our samples.  Of the emission lines to which we can fit two
components (\lya, \civ, \mgii, and \hb), it is the {\em only} line for
which the percentage of total line flux in the narrow line component
may be {\em larger} in the XF composite.  Since \hb\, appears to be unique
among the larger variety of emission lines studied here, it is clear
that \hb\, may not be the best, or only representative of BLR line widths.
Fitting the \hb\, line in our optical composites with a
{\em single} Gaussian component leads to similar conclusions to these previous
studies (i.e., that FWHM(\hb) is larger in the XF composite).  
However, fitting narrow and broad line components separately reveals
that \hb\, has a stronger broad component in the XB composite, and a
larger broad/narrow line ratio, while the actual FWHM is similar to
the XF composite.  This reveals that single Gaussian fits to compound
lines must be interpreted with caution.   Furthermore, we also 
caution that the luminosity ranges for these (PG and LBQS) samples 
are disjoint. However, since \ax\, and not \aox\, seems to have the
more fundamental correlation with both \hb\, and \oiii\, in Laor et
al. (1997), we highlight the need for correlation of \ax\, with a
wider range of emission line measurements.  

The parameters emphasized here, \aox\, and narrow line emission, also
appear to be linked to the following quantities: X-ray spectral slope
\ax, \feii\, strength, luminosity, radio loudness (e.g., Laor et
al. 1997, Lawrence et al. 1997, Green et al. 1995, BG92).  We suggest
here that the as yet mysterious physical link between these diverse
properties is intimately related to high velocity, outflowing winds
near the nucleus that, by absorbing the intrinsic nuclear continuum,
strongly affect the radiation observed at larger distances.  The
continuum impinging on the NLR is closest to that received by distant
observers like ourselves, but is quite different from that arriving at
the BLR.  Material in the BLR itself may reprocess the intrinsic
continuum issuing from the unshrouded QSO nucleus. 

Only some of the correlations between measured
line and continuum parameters are intrinsic, and others
simply add dispersion to more primary correlations. The principal 
physical processes must be extracted from the principle observational
eigenvectors in a multivariate, multiwavelength approach,
with careful attention to continuum slopes and detailed emission
line fits.   We are accumulating a high-quality homogeneous database
including all this information for a large sample of QSOs, primarily
from the ROSAT and HST archives.  We believe that multiwavelength
studies such as this show promise for significant advances in our
understanding.

Thanks to Marianne Vestergaard for performing the iron subtraction,
to Craig Foltz for the LBQS spectra, Ken Lanzetta for the
IUE QSO atlas, and Todd Boroson for the optical \feii\, template.
The author gratefully acknowledges support provided by NASA
through Grant NAG5-1253, and Contract NAS8-39073 (ASC), as well as
HF-1032.01-92A awarded by the Space Telescope Science Institute, which
is operated by the Association of Universities for Research in
Astronomy, Inc., under NASA contract NAS5-26555.  This research has
made use of the NASA/IPAC Extragalactic Database (NED) which is
operated by the Jet Propulsion Laboratory, California Institute of
Technology, under contract with the National Aeronautics and Space
Administration.

\end{document}


\begin{tabular}{llcccccrl} 
\multicolumn{9}{c}{ TABLE 1 } \\
\multicolumn{9}{c}{ CONTINUUM PARAMETERS FOR QSO SUBSAMPLES} \\
\hline\hline  
\multicolumn{9}{c}{ } \\
Parameter & Subsample & \multicolumn{2}{c}{ Sample Size}  & Median  &
Mean & RMS & Trunc?$^b$ & Prob (\%)$^c$  \\ 
     &     & N$_{total}$ & N$_{limits}^a$  &      &      &       &
&  \\
\hline
redshift & LBQS/XF & 54 & 0 & 0.46 & 0.49 & 0.03 & N & 43    \\
         & LBQS/XB & 60 & 0 & 0.45 & 0.51 & 0.03 & N &     \\
         & IUE/XF  & 23 & 0 & 0.35 & 0.57 & 0.09 & N & 2.5    \\
         & IUE/XB  & 25 & 0 & 0.21 & 0.34 & 0.03 & N &     \\
\multicolumn{9}{c}{} \\
\loglopt & LBQS/XF & 54 & 0 & 30.41 & 30.44 & 0.06 & N & 17    \\
         & LBQS/XB & 60 & 0 & 30.30 & 30.30 & 0.06 & N &     \\
         & IUE/XF  & 23 & 0 & 31.03 & 31.05 & 0.15 & N & 25    \\
         & IUE/XB  & 25 & 0 & 30.50 & 30.57 & 0.08 & N &     \\
\multicolumn{9}{c}{} \\
\aox     & LBQS/XF & 54 & 42 & 1.61 & 1.61 & 0.01 & 1.67 & $<10^{-4}$  \\
         & LBQS/XB & 60 & 0 &  1.34 & 1.35 & 0.01 & N &     \\
         & IUE/XF  & 23 & 4 & 1.57 & 1.60 & 0.03 & N & $<10^{-4}$  \\
         & IUE/XB  & 25 & 0 & 1.31 & 1.28 & 0.02 & N &   \\
\multicolumn{9}{c}{} \\
\loglx   & LBQS/XF & 54 & 42 & 22.55 & 25.96 & 0.08 & 25.64 & $<10^{-4}$  \\
         & LBQS/XB & 60 & 0 & 26.78 & 26.78 & 0.06 & N &     \\
         & IUE/XF  & 23 & 0 & 26.91 & 26.78 & 0.20 & N & 1    \\
         & IUE/XB  & 25 & 0 & 27.17 & 27.25 & 0.09 & N &     \\
\hline
\end{tabular}\\[0.5ex]

\parbox[b]{5in}{
NOTE.~-- ~ For LBQS/XF: $z<1$; $\aox \ge 1.475$; 
for LBQS/XB: $z<1$; $\aox < 1.475$; detections only;
for IUE/XF: $\aox \ge 1.4$; for IUE/XB: $\aox < 1.4$; detections only.

\hskip0.25in $^a$~All limits are upper limits except for \aox. 

\hskip0.25in $^b$~ This column lists the value of an outlying 
non-detection, if it was necessarily redefined, truncating 
the distribution to allow normalization.  The resulting mean value is
biased.  

\hskip0.25in $^c$~ Probability that the two subsample distributions
for this parameter are randomly drawn from the same parent
distribution, from ASURV logrank test. 
}

\clearpage

\begin{tabular}{ccccc} 
\multicolumn{5}{c}{ TABLE 2 } \\
\multicolumn{5}{c}{ COMPOSITE NORMALIZATION DEFINITIONS } \\
\hline\hline  
\multicolumn{5}{c}{ } \\
Wavelength & $\lambda_n$ &  Continuum  & Continuum  & Line Name  \\
Region     &             &  Band       & Band       &        \\
\hline
\multicolumn{5}{c}{ } \\
\multicolumn{5}{c}{LBQS Spectra} \\
4150-5100 & 4863 & 4710 - 4780 & 5040 - 5090 & \hb    \nl
3800-4800 & 4102 & 4020 - 4050 & 4150 - 4270 & \hd    \nl
3200-4000 & 3426 & 3360 - 3400 & 3450 - 3500 & \nev   \nl
2000-3000 & 2798 & 2645 - 2700 & 3020 - 3100 & \mgii  \nl
\multicolumn{5}{c}{} \\
\multicolumn{5}{c}{IUE Spectra} \\
1000-2000 & 1549 & 1450 - 1480 & 1680 - 1700 & \civ \nl
\hline
\end{tabular}\\[0.5ex]

\parbox[b]{9cm}{

\hskip0.2cm
NOTE.~--~  
Each individual rest-frame QSO spectrum is divided by the flux of the
fit line at the normalization wavelength $\lambda_n$ before
addition to a composite spectrum.  The continuum level at $\lambda_n$ 
is found by fitting a linear continuum between two points; the mean
flux values in the two continuum bands shown, at their central
wavelength points. 

}

\clearpage

\begin{tabular}{lcrr} 
\multicolumn{4}{c}{ TABLE 3 } \\
\multicolumn{4}{c}{ SPECFIT RESULTS FOR LBQS COMPOSITES} \\
\hline\hline  
\multicolumn{4}{c}{ } \\
Component & Parameter & XB Composite  & XF Composite  \\
\hline
\multicolumn{4}{c}{} \\
2000-3000\AA & f$_{1000}$ &  2.44 $\pm$   0.07 &     1.79 $\pm$   0.04 \\
continuum  & $\alpha$ &      0.89 $\pm$   0.03 &     0.59 $\pm$   0.02  \\
\multicolumn{4}{c}{} \\
\feii\, 
 &f$~^a$ &      41.1 $\pm$  0.2   &   36.8 $\pm$   0.1  \\
\multicolumn{4}{c}{} \\
\mgii\, broad$^b$
 &f &      42.0 $\pm$   2.5 &    36.4 $\pm$   2.1  \\
 & $\lambda_c$ &    2796.9 $\pm$   3.5 &  2801.6$^c$   \\
\multicolumn{4}{c}{} \\
\mgii\, narrow 
 &f &      15.94 $\pm$   1.38 &    11.26 $\pm$   1.11  \\
 & $\lambda_c$ &    2798.8 $\pm$   0.8 &  2801.6 $\pm$   0.8  \\
 & FWHM &    2650 $\pm$ 230 &  2530 $\pm$ 240  \\
\multicolumn{4}{c}{} \\
3200-4000\AA & f$_{1000}$ &        31.3 $\pm$   1.2 &    29.2 $\pm$   0.8  \\
continuum  & $\alpha$ &       2.81 $\pm$   0.03 &     2.76 $\pm$   0.02  \\
\multicolumn{4}{c}{} \\
NeV$^d$ 
 &f &       1.89 $\pm$   0.22 &     1.06 $\pm$   0.15  \\
 & $\lambda_c$ &    3425.6 $\pm$   1.2 &  3425.5 $\pm$   1.8  \\
\multicolumn{4}{c}{} \\
\oii$^e$ 
 &f &       2.10 $\pm$   0.24 &     0.65 $\pm$   0.17  \\
 & FWHM &     900 $\pm$ 120 &  1090 $\pm$ 290  \\
\multicolumn{4}{c}{} \\
3800-4800\AA & f$_{1000}$ &   13.5 $\pm$   1.3 &    35.6 $\pm$   3.1  \\
continuum & $\alpha$ &      1.82 $\pm$   0.07 &     2.50 $\pm$   0.06  \\
\multicolumn{4}{c}{} \\
\neiii 
 &f &       2.9 $\pm$   1.2 &     2.3 $\pm$   0.6  \\
 & $\lambda_c$ &    3870.8 $\pm$   3.6 &  3867.3 $\pm$   4.1  \\
 & FWHM &    1330 $\pm$ 240 &  1200$^f$  \\
\multicolumn{4}{c}{} \\
\hei 
 &f &       1.45 $\pm$   1.3 &     1.43 $\pm$   0.6  \\
 & $\lambda_c$ &    3894 $\pm$  15 &  3887 $\pm$   6  \\
 & FWHM &    2100 $\pm$ 1500 &  1200$^f$   \\
\multicolumn{4}{c}{} \\
\sii$+$\hd 
 &f &       4.06 $\pm$   0.6 &     4.10 $\pm$   1.0  \\
 & $\lambda_c$ &    4103.0 $\pm$   1.5 &  4106.7 $\pm$   6.1  \\
 & FWHM &    1330 $\pm$ 240 &  3440 $\pm$ 780  \\
\multicolumn{4}{c}{} \\
\hg$+$\oiii 
 &f &      15.3 $\pm$   1.0 &    20.7 $\pm$   1.7  \\
 & $\lambda_c$ &    4347.5 $\pm$   1.5 &  4348.4 $\pm$   2.8  \\
 & FWHM &    3240 $\pm$ 250 &  5200 $\pm$ 620  \\
\multicolumn{4}{c}{} \\
\hline
\end{tabular}\\[0.5ex]

\begin{tabular}{lcrr} 
\multicolumn{4}{c}{ TABLE 3 - continued } \\
\multicolumn{4}{c}{ SPECFIT RESULTS FOR LBQS COMPOSITES} \\
\hline\hline  
\multicolumn{4}{c}{ } \\
Component & Parameter & XB Composite  & XF Composite  \\
\hline
\multicolumn{4}{c}{} \\
4150-5100 & f$_{1000}$ &       30.0 $\pm$   3.3 &    47.0 $\pm$   3.4 \\
continuum & $\alpha$ &      2.17 $\pm$   0.07 &     2.43 $\pm$   0.05  \\
\multicolumn{4}{c}{} \\
\feii\, 
 &f$~^g$ &    26.5 $\pm$  0.2   &   31.1 $\pm$  0.1  \\
\multicolumn{4}{c}{} \\
\hb\, Broad 
 &f &    64.0 $\pm$   2.6   &   42.8 $\pm$   2.5  \\
 & $\lambda_c$ &  4869.3 $\pm$   1.7 &   4872.5 $\pm$   2.5  \\
 & FWHM &  5660 $\pm$ 290  &  5730 $\pm$ 360  \\
\multicolumn{4}{c}{} \\
\hb\, narrow
 &f &     9.1 $\pm$   1.4   &   10.7 $\pm$   1.7  \\
 & $\lambda_c$ &  4865.6 $\pm$   1.1 &   4864.2 $\pm$   1.3  \\
 & FWHM$^h$ &   976  &   1330   \\
\multicolumn{4}{c}{} \\
$\oiii\lambda 5007$
 &f &    32.32 $\pm$   1.11  &    13.46$^i$ $\pm$   0.95  \\
 & $\lambda_c$ &  5010.2 $\pm$   0.3 &   5009.9 $\pm$   0.8  \\
 & FWHM &   976 $\pm$  43  &  1330 $\pm$ 108  \\
\multicolumn{4}{c}{} \\
\hline
\end{tabular}\\[0.5ex]

\parbox[b]{4in}{

\hskip0.2cm
NOTE.~--~  
All fluxes are summed over a locally normalized continuum (see text) 
and so are approximately equivalent widths.  The value of the
(normalized) continuum flux at any wavelength $\lambda$ can be
obtained via f$_c = {\rm f}_{1000}(\lambda/1000)^{-\alpha}$, Emission line
centroids $\lambda_c$ are in \AA, while FWHM are shown in \kms.

\hskip0.2cm $^a$ \feii\, summed over PL fit from $\lambda\lambda
2210-2730$.

\hskip0.2cm $^b$ FWHM fixed at 15000~\kms.

\hskip0.2cm $^c$ $\lambda_c$ fixed at narrow \mgii\, values.

\hskip0.2cm $^d$ FWHM fixed at 1000~\kms.

\hskip0.2cm $^e$ $\lambda_c$ fixed at 3727\AA.

\hskip0.2cm $^f$ FWHM fixed at 1200~\kms.

\hskip0.2cm $^g$ \feii\, summed over PL fit from $\lambda\lambda
4434-4684$.

\hskip0.2cm $^h$ FWHM fixed at $\oiii\lambda 5007$ FWHM.

\hskip0.2cm $^i$ \woiii\, of XF composite decreases to 
$7.5\pm1$ after  \feii\, subtraction.
}

\clearpage
\clearpage

\begin{tabular}{lcrr} 
\multicolumn{4}{c}{ TABLE 4 } \\
\multicolumn{4}{c}{ SPECFIT RESULTS FOR IUE COMPOSITES} \\
\hline\hline  
\multicolumn{4}{c}{ } \\
Component & Parameter & XB Composite  & XF Composite  \\
\hline
\multicolumn{4}{c}{} \\
1000-2000\AA  &
 $\alpha$ &  1.51 $\pm$  0.02 &  1.57 $\pm$  0.02 \\ 
continuum &  f$_{1000}$ & 0.75 $\pm$  0.02 &      0.75 $\pm$  0.03 \\ 
\multicolumn{4}{c}{}  \\
\lyb/\ovi\,
 &  f &     19.6 $\pm$  1.1 &     10.1 $\pm$  1.0 \\ 
 &  $\lambda_c$ &   1033.5 $\pm$  0.4 &   1032.5 $\pm$  0.6 \\ 
 &  FWHM &   5230 $\pm$ 400 &   4560 $\pm$ 450 \\ 
\multicolumn{4}{c}{}  \\
\lya\, broad  
 & f  &     87.3 $\pm$ 3.9 &  78.7 $\pm$ 2.9 \\ 
 & $\lambda_c$ &   1218.8 $\pm$  0.8 &   1218.7 $\pm$  0.5 \\ 
 & FWHM &  12840 $\pm$ 410 &  12700 $\pm$ 400 \\ 
\multicolumn{4}{c}{}  \\
\lya\, narrow  
 & f  &     47.3 $\pm$  1.7 &  36.8 $\pm$  2.2 \\ 
 & $\lambda_c$ &   1214.6 $\pm$  0.1 &   1214.9 $\pm$  0.2 \\ 
 & FWHM &   3030 $\pm$ 80 &   3470 $\pm$ 130 \\ 
\multicolumn{4}{c}{}  \\
NV  
 & f  &  3.9 $\pm$  2.5 &   1.2 $\pm$  0.3 \\ 
 & $\lambda_c$ &   1240 $\pm$  3 &   1244 $\pm$  1 \\ 
 & FWHM &   6000 $\pm$ 1200 &   6700 $\pm$ 3200 \\ 
\multicolumn{4}{c}{}  \\
OI 
 & f  &   2.6 $\pm$  0.9 &   6.8 $\pm$  1.9 \\ 
 & $\lambda_c$ &   1310 $\pm$  4 &   1312 $\pm$  4 \\ 
 & FWHM &   6700 $\pm$ 3400 &   6400 $\pm$ 2100 \\ 
\multicolumn{4}{c}{}  \\
\siiv\,
 & f  &   8.4 $\pm$  1.0 &   11.2 $\pm$  1.6 \\ 
 & $\lambda_c$ &   1398 $\pm$  1.7 &   1395 $\pm$  2.3 \\ 
 & FWHM &   6250 $\pm$ 870 &   6270 $\pm$ 950 \\ 
\multicolumn{4}{c}{}  \\
\civ\,  broad  
 & f  &     52.9 $\pm$  1.3 &     39.2 $\pm$  9.0 \\ 
 & $\lambda_c$ &   1546.2 $\pm$  1.6 &   1548.3 $\pm$  1.6 \\ 
 & FWHM &  12500 $\pm$ 2800 &  12000 $\pm$ 2500 \\ 
\multicolumn{4}{c}{}  \\
\civ\, narrow
 & f  &  30.6 $\pm$  6.6 &  14.8 $\pm$  7.9 \\ 
 & $\lambda_c$ &   1549.09 $\pm$  0.2 &   1548.34 $\pm$  0.6 \\ 
 & FWHM &   3200 $\pm$ 350 &   3400 $\pm$ 940 \\ 
\multicolumn{4}{c}{}  \\
absorption
 & f  &      5.8 $\pm$  3.9 &   8.8 $\pm$  1.4 \\ 
 & $\lambda_c$ &   1487 $\pm$ 19 &   1480 $\pm$ 12 \\ 
 & FWHM &  12900 $\pm$ 1800 &  12500 $\pm$ 2800 \\ 
\multicolumn{4}{c}{}  \\
\hline
\end{tabular}\\[0.5ex]
\parbox[b]{4in}{

\hskip0.2cm
NOTE.~--~  Parameters have the same meaning as in Table~3. 

}